\documentstyle[prl,aps,psfig]{revtex}
\twocolumn
\begin{document}
\draft
\twocolumn[\hsize\textwidth\columnwidth\hsize\csname@twocolumnfalse\endcsname
\title{Velocity Fluctuations in 
Electrostatically Driven Granular Media}
\author{I. S. Aranson$^{1}$ and  J. S. Olafsen$^{2}$} 
\address
{$^{1}$Argonne National Laboratory, 9700 S. Cass Avenue, Argonne, IL 60439 \\
$^{2}$Department of Physics and Astronomy, University of Kansas, Lawrence, 
KS 66045 }
\maketitle
\begin{abstract}
We study experimentally the particle velocity fluctuations 
in an electrostatically driven dilute granular gas. The
velocity distribution functions have strong deviations
from Maxwellian form  in a wide range of parameters. We have found that 
the tails of the distribution functions are consistent with a stretched 
exponential law with typical exponents of the order $3/2$. 
Molecular dynamic simulations shows qualitative agreement with experimental data. 
Our results suggest that this non-Gaussian behavior is typical for 
most inelastic gases with both short and long range interactions.  
\end{abstract}
\vskip1pc]
\narrowtext

Despite extensive study over the preceding decade,
a fundamental understanding of the dynamics of
granular materials still poses
a challenge for
physicists and engineers \cite{jnb,kadanoff}.  
Driven granular materials exhibit complex behavior
that resemble some aspects of conventional
solids, liquids, and gases,
yet also demonstrate
some considerable differences\cite{jnb,kadanoff,gennes,raj}.
%A key distinction that separates granular phases from their
%conventional counterparts is the presence of
%dissipation of energy through inelasticity of collisions  and friction.

Recent experimental studies  with vibrationally--driven 
granular gases revealed surprising deviations in the particle 
distribution function from the Maxwell 
distribution law \cite{menon,olafsen,olafsen1,kudrolli,losert,blair1}. These 
deviations were attributed to  the effects of dissipation due to 
inelasticity of inter-particle collisions. In particular, 
Ref. \cite{menon} reported an ``universal'' stretched exponential law 
\begin{equation} 
P(v)  \sim \exp[-|v/v_0|^\zeta] 
\label{zeta} 
\end{equation} 
with the exponent $\zeta=3/2$. 
Here $v$ is particle velocity, 
$P$ is the velocity distribution function and $
v_0$ is the ``thermal velocity''. 
This behavior was observed in a wide 
range of frequencies and amplitudes of vibration and for different 
densities of the granular gas and 
agrees with 
the theoretical prediction of Ref. \cite{ernst}.  
Deviations from 
Maxwellian  behavior also has been observed experimentally in
different geometries and for different driving conditions in Refs. 
\cite{olafsen1,kudrolli,losert} and in 
numerical
simulations \cite{carrillo,shattuck,soto}.
These studies 
suggest that the deviations from a Maxwell distribution are the result of 
short-range inelastic hard-core collisions between particles. 

Interactions between particles often  are 
not reduced to simple hard-core collisions.
Fascinating collective behavior appears
when small particles acquire an electric charge and respond to
competing long-range electromagnetic
and short range contact forces. An important question is whether 
non-Maxwellian distributions are typical only for granular  gases 
with hard-core collisions or could it also be observed for general dissipative 
gases with both short-range and long-range interactions.

The electrostatic excitation of
granular media  offers unique new
opportunities compared
to traditional  vibration techniques which have been developed
to explore granular dynamics.
It enables one to deal with
extremely fine powders which are not easily controlled
by mechanical methods.
Electrostatic driving makes use of these {\it bulk} forces, and  allows
control of the ratio between long-range electric forces and 
short-range collisions by changing the amplitude and the frequency 
of the applied electric field. Our previous studies with 
electrostatically-driven granular media revealed
a hysteretic phase transition from the immobile
condensed state (granular solid)
to a fluidized dilated  state
(granular gas) with a changing applied electric field \cite{blair}.
A spontaneous precipitation of  dense
clusters from the gas phase and a subsequent coarsening -- coagulation of
these clusters -- is observed in a certain region of the electric 
field values. The strong effect of humidity 
on dynamics of electrostatically-driven
granular materials was studied in Ref. \cite{howell}.  
%It was found the clustering dynamics in conducting particles is
%primarily controlled by screening of the electric field
%but is aided by cohesion due to humidity.
%It is shown that humidity effects
%dominate the clustering process with dielectric particles.

In this Letter we study  the particle velocity 
distributions in electrostatically driven granular media.
We have found that in a wide range of parameters 
the particle velocity distribution function is strongly
non-Maxwellian and is well-approximated by the stretched 
exponential law $P(v) \sim \exp (-|v/v_0|^{3/2})$. 
We performed molecular dynamics simulations of conducting particles in 
an ac electric field and have obtained qualitative agreement with the experiment. 
We conclude from our results that 
the tails of the velocity distributions for 
driven granular gases in general exhibit non-Maxwellian behavior and are not
limited to the systems with just hard-core collisions.

\begin{figure}
\centerline{ \psfig{figure=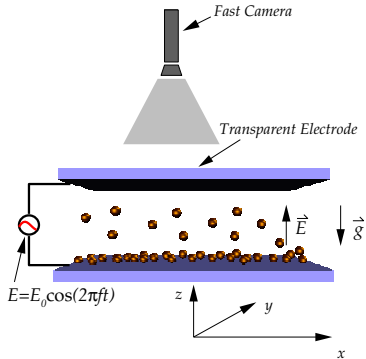,height=2in}}
\caption{Block-diagram of experimental apparatus. 
}
\label{Fig1}
\end{figure}

Our experimental setup is similar to that in Ref. \cite{blair,howell}. 
Particles are placed between
the plates of a large capacitor which is energized by
a constant (dc) or alternating  (ac) electric field $E=E_0 \cos(\omega t)$,
see Fig. \ref{Fig1}.
To provide optical access to  the cell, the capacitor 
plates were  made of glass with a clear conductive coating.
We used $11\times11$ cm capacitor plates
with a spacing of 1.5 mm.  The particles consisted of
165 $\mu m$ conducting bronze spheres. 
The field amplitude $E_0$ varied from 0 to 10 kV/cm
and the frequencies $f=\omega/2 \pi$ on the interval of 0 to 120 Hz.
The total number of particles in the cell varied between $10^5$ and $10^6$.

Conducting particles
acquire a surface
charge when they are
in contact with the
capacitor plate.
As the magnitude of the electric field
in the capacitor exceeds a
critical value, $E_1$, the upward electric
force overcomes gravity $mg$ ($m$ is the mass of the particle,
$g$ is the gravity)
and lifts  the charged particles.
When  grains hit the upper plate, they deposit  their charge and fall
back.
By applying an alternating electric field
$E=E_0 \sin (2 \pi   t)$, and adjusting its frequency $f$,
one can control the vertical excursion of particles by effectively
turning them back before they collide with the upper plate.
Thus, by increasing the frequency of the electric field, the particles 
can be confined in a relatively thin layer, i.e.  the 
granular  gas is quasi-two-dimensional at high frequencies and 
three-dimensional at low frequencies. It effectively allows for  
control over the number of collisions and, therefore, the contributions from  long-range 
and short-range interactions. 
This control over the vertical extent of the motion of the particles also
changes the manner in which energy is transferred from the vertical to the 
horizontal through collisions. From the comparison of typical kinetic energy 
acquired by the particle and electric energy of inter-particle interaction 
it is possible to show that the long-range electric interaction becomes dominant 
with respect to short-range collisions at the frequencies larger than $f_0
\approx  100$ Hz for given particle size. 
%Also, at lower  frequencies for fixed driving amplitude,
%there is a trend toward larger horizontal granular temperatures, $T_g = \langle v^2 
%\rangle $.  

In our experiment we extracted horizontal particles velocities via high-speed image analysis. 
Pictures were obtained at the  rate  1000 frames per second from a camera
mounted to a microscope suspended vertically above the cell and 
particles positions were resolved to sub-pixel resolution. Inter-particle
and particle-boundary collisions that introduce sudden changes in momenta
of the particles were filtered from the distributions in a manner similar 
to Ref \cite{olafsen,olafsen1}.  These events were only a small fraction of the total
number of measurements due to the low density at which they were acquired.
An ensemble average for each of the velocity distributions were obtained 
with nearly $10^6$ data points.

\begin{figure}
\centerline{ \psfig{figure=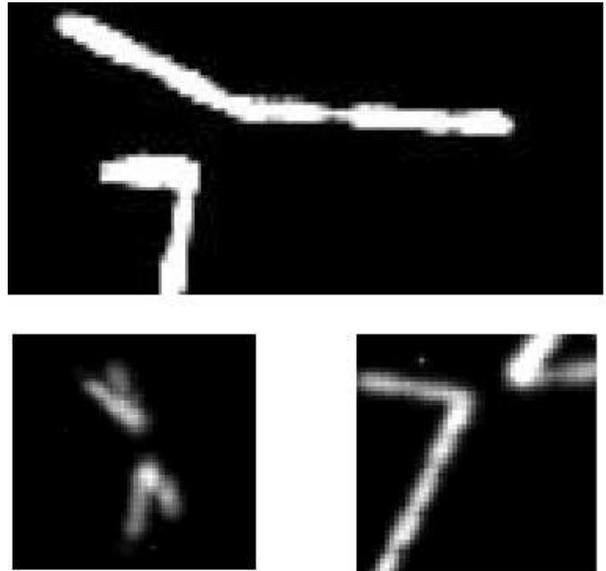,height=3in}}
\caption{Composite images of  two particles collision   
created by overlaying
several sequential images taken with the high speed camera for a driving
frequency of $50Hz$.  Each of the three  images  
demonstrates a different collision between two particles that occurs at a distance
larger than a ball diameter.  At this relatively high frequency the particles 
have a small
vertical excursion and the focal plane of the microscope is set between the two plates
of the cell.  The particles are completely illuminated which results in the smudged streaks 
of varying intensity in the particle tracks. 
}
\label{Figcol}
\end{figure}

The effect of long-range electric forces is illustrated in Fig. \ref{Figcol}. 
In contrast to vibrationally-driven systems, the momentum transfer 
occurs often  without actual collision. While there are hard sphere
collisions in the system as well, this interaction occurs at a center
to center distance of approximately two particle diameters.

\begin{figure}
\centerline{ \psfig{figure=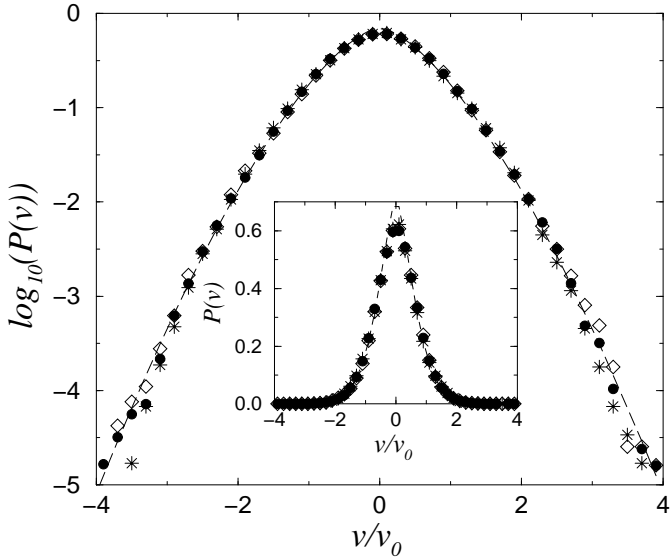,height=3in}}
\caption{Main plot: The logarithms of horizontal velocity distribution $P(v)$ 
vs normalized velocity $v/v_0$, where $v_0=\sqrt{2 T_g}$, $T_g$ is horizontal 
granular temperature. Bullets correspond to $f=50$ Hz and $E_0=8.83$ kV/cm, 
stars to $f=120$ Hz and $E_0=8.83$ kV/cm, diamonds to $f=45$ Hz and $E_0=7.3$ kV/cm.
Dashed line show the best fit according to Eq. (\ref{zeta}) with $\zeta\approx1.51$. 
Inset:  The horizontal velocity distribution $P(v)$
vs normalized velocity $v/v_0$ in linear scale. 
}
\label{FigExp}
\end{figure}

\begin{figure}
\centerline{ \psfig{figure=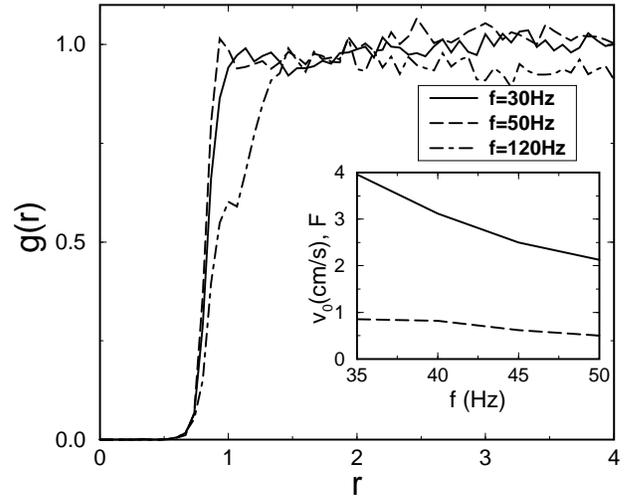,height=3in}}
\caption{ 
The density distribution function $g(r)$ for different values of $f$. 
Inset: ``thermal velocity'' $v_0$ (solid line) and flatness $F$ 
(dashed line)  vs $f$  for $E_0=8.83$
kV/cm.
}
\label{FigExp0}
\end{figure}

A summary of experimental results is presented in Figs. \ref{FigExp} 
and \ref{FigExp0}. 
We find for various values of frequency $f$ and electric field amplitude 
$E_0$ the normalized velocity distributions both in logarithmic and 
linear scales lay practically on a top of each other, 
see Fig. \ref{FigExp}. There is a very small difference in granular 
temperature $T_g$ between $x$ and $y$ directions, which is probably 
related to some small tilt of the cell with respect to gravity.
There is an increase in the flatness of the velocity distributions, 
$F=\langle v^4 \rangle /\langle v^2 \rangle ^2 - 3$, 
as the driving frequency is decreased at a fixed driving amplitude, 
see Fig. \ref{FigExp0},inset. (As defined here,
a flatness of 0 corresponds to a Gaussian distribution). 
In a mechanically shaken 
granular layer, an increase in flatness was observed as the 
shaking amplitude was decreased at 
constant frequency, which corresponded to a lower
horizontal granular temperature \cite{olafsen1}.  
In the electrostatically driven system, the increase in 
flatness occurs as the horizontal granular
temperature increases as there is more kinetic energy to 
transfer from the horizontal to the
vertical direction because of the increase in
vertical excursion at lower frequencies.     

Fig. \ref{FigExp0} shows the density distribution function 
$g(r)=\langle \rho(r)/\rho(0)\rangle /\langle \rho \rangle ^2$ 
for different values of frequency. 
The volume density for the experimental conditions 
is extremely dilute ($\approx 7\%$) 
as can be seen by the lack of particle correlations with the
exception of the excluded volume of the finite particles.
Consequently,  
the function $g(r)$ has almost step-like form for low frequencies. 
One can also see that with the increase of  frequency
the contribution from long-range interactions also increases. 
This fact is manifested in the smearing of the step-like 
structure of $g(r)$, which can be interpreted as an 
increase of the effective crossection for particle interactions.
 
In order to obtain an insight into the problem we performed molecular dynamics 
simulations of conducting particles in an applied ac electric field. We modified 
the MD code used in our earlier paper \cite{blair}. In particular, we introduced a finite 
roughness of the bottom plate and finite 
restitution coefficient $r$ between the particles and particles and 
walls ($r=0.8$ and $r=0.6$ correspondingly) \cite{footnote1}. We changed the restitution 
coefficient and roughness over a wide range, but the  results  appear to
be very robust to the specific choice of $r$.  
We simulated up to 1000 particles 
in the cell $80\times80$ particles diameters in $x-y$ plane, electric field 
amplitude $E_0$ was $2.6 E_1$, where $E_1$ is the first critical field value 
(for $E=E_1$ upward eclectic force  acting on isolated particle overcomes gravity, 
see \cite{blair}). The velocity is scaled by $\bar v =\sqrt{g a}$ and time by $\bar t =\sqrt{a/g}$, 
where $a$ is the radius of particle. Typical simulation time was 10,000 $\bar t $, and the velocity 
distribution was built from 2,500,000 sample points. 

\begin{figure}
\centerline{ \psfig{figure=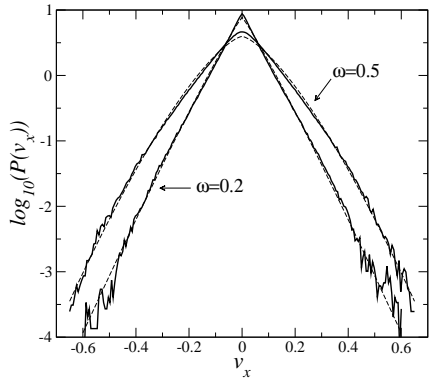,height=3in}}
\caption{
The logarithms of horizontal velocity distribution $P(v_x)$  ($x$-component only) for two
different values of frequency $\omega$. Solid lines show numerical results, 
dashed lines are fits $\log P(v)=a_0-a_1 v^\zeta$, with $\zeta \approx 1.43$ for $\omega=0.5$
and $\zeta \approx 1.08$ for $\omega=0.2$. 
}
\label{FigMD}
\end{figure}

Selected results are presented in Fig. \ref{FigMD}. 
In the wide range of parameters $\omega$, $E_0$ we have found non-Maxwellian velocity 
distributions. For relatively high frequencies $\omega=0.5$ (corresponding approximately 
to $f=30$ Hz for our experimental parameters) the velocity distribution is indeed 
approximated by Eq. (\ref{zeta}) with $\zeta$ close to $3/2$,  as in the experiment. 
For a smaller frequency ($\omega=0.2$) we have found that 
$P(v)$ is consistent with a simple exponential 
law: $P(v) \sim \exp(-v/v_0)$, i.e. $\zeta = 1$. 
It is plausible that the crossover to exponent $\zeta=3/2$ occurs for 
the high velocity tail and would explain the gentle increase in flatness as the 
frequency is reduced.
However, one likely needs substantially longer simulations in order 
to resolve this crossover.

Our simulation results for ``high frequency'' yield an exponent $\zeta\approx1.5$ 
and  are in a good agreement with 
experiment. Unfortunately, we cannot verify the ``low frequency'' exponent 
$\zeta=1$ because our current experimental apparatus does not allow us to 
extract reliable particles positions for the frequencies  $f$ below $30$ Hz. 
For small frequencies the vertical particle displacements are much higher 
than the particle size. As a result, the particles go out of the focal plane
of the microscope and it is impossible to sufficiently resolve horizontal positions.

Let us briefly discuss the results. Universal high velocity tails of the distribution 
$P \exp(-|v/v_0|^{3/2})$ were derived by van Noije and Ernst \cite{ernst} 
from the Enskog-Boltzmann equation
for  the uniformly heated state of a 2D weakly-inelastic granular gas.  
According to \cite{ernst}, in the steady state, the averaged distribution 
function satisfies the equation 
\begin{equation} 
D \partial ^2_v  P = - I(P,P) 
\end{equation} 
where $I(P,P)$ is the pair collision integral and $D$ is proportional to 
the strength of external ``thermal noise''. In the large velocity limit 
Ref. \cite{ernst} derived $I(P,P) \sim v P$ since the fast particle 
typically collides with particles in the ``thermal range'',  which immediately results in 
Eq. (\ref{zeta})  for the high velocity tail. 

Although there is no ``thermal noise'' in our experiment, multiple collisions between 
particles,  the finite roughness of the plates, 
and long-range interactions result in sufficient randomization of particle velocities 
and presumably have an overall effect similar to thermal noise. The argument  
that the collision integral  for high velocities is reduced to $v P$ appears to be 
very robust and does not depend on the specifics of the model. Thus, we can expect 
the above stretched exponential velocity tails to be characteristic of our system as well. 

The distribution $P \sim \exp(-v/v_0)$ was derived 
for a freely evolving granular gas \cite{esipov}. One can speculate that  our results for 
the small frequency  have some fingerprints of a ``freely evolving gas''
because the  particles spend significant time in  the flights 
between plates and exhibit some sort of ``free cooling'', whereas for the higher frequencies
the particles are confined in a 
thin layer and collisions with the bottom plate occur more often.

In conclusion, 
we experimentally and numerically studied particle velocity
distributions in an electrostatically driven granular media.
We have found that in a wide range of experimental parameters
the particle velocity distribution function is strongly
non-Maxwellian and is well-approximated by the stretched
exponential law $P(v) \sim \exp (-|v/v_0|^{3/2})$.  
Our results suggest that such velocity distributions should be generic 
for dissipative gases with both short and long range interactions  and 
may possibly have some relevance for completely different systems, 
such as dusty plasmas, colloids, etc.

We thank B. Meerson for valuable comments. 
This research was supported by the US DOE, Office of Basic Energy Sciences, 
contract W-31-109-ENG-38. 
JSO would like to acknowledge the donors of the Petroleum Research
Fund, administered by the ACS, for partial support of this research as well as the
support of the National Science Foundation under Grant No. EPS-9874732 and matching
support from the State of Kansas.

\end{document}